\documentclass[useAMS,usenatbib]{mnras}

\usepackage[utf8]{inputenc}
\usepackage{graphicx}   
\usepackage{amsmath}    
\usepackage{amssymb}    
\usepackage{multicol}   
\usepackage{bm}         
\usepackage{pdflscape}  
\usepackage{float}      
\usepackage{tipa}       
\usepackage{orcidlink}
\usepackage{color, soul}
\soulregister\cite7
\soulregister\ref7
\soulregister\pageref7

\hypersetup{final}
\title[]{Heating and cooling in stellar coronae: coronal rain on a young Sun}
\author[]{Daley-Yates S.}

\author[Simon Daley-Yates, Moira M. Jardine and Craig D. Johnston]{
    S. Daley-Yates$^{1}$\thanks{E-mail: sddy1@st-andrews.ac.uk} \orcidlink{0000-0002-0461-3029},
    Moira M. Jardine$^{1}$ \orcidlink{0000-0002-1466-5236} and
    Craig D. Johnston$^{2,3}$ \orcidlink{0000-0003-4023-9887}\\   
    $^{1}$School of Physics and Astronomy, University of St Andrews, North Haugh, St Andrews, Fife, Scotland KY16 YSS, UK\\
    $^{2}$Department of Physics and Astronomy, George Mason University, Fairfax, VA 22030, USA\\
    $^{3}$Heliophysics Science Division, NASA Goddard Space Flight Center, Greenbelt, MD 20771, USA
}

\begin{document}

\date{}

\pagerange{\pageref{firstpage}--\pageref{lastpage}} \pubyear{2023}

\maketitle

\label{firstpage}

\begin{abstract}
Recent observations of rapidly-rotating cool dwarfs have revealed H$\alpha$ line asymmetries indicative of clumps of cool, dense plasma in the stars' coronae. These clumps may be either long-lived (persisting for more than one stellar rotation) or dynamic. The fastest dynamic features show velocities greater than the escape speed, suggesting that they may be centrifugally ejected from the star, contributing to the stellar angular momentum loss. Many however show lower velocities, similar to coronal rain observed on the Sun. We present 2.5D magnetohydrodynamic simulations of the formation and dynamics of these condensations in a rapidly rotating ($P_{\rm rot}~=~ 1 \ \mathrm{day}$) young Sun. Formation is triggered by excess surface heating. This pushes the system out of thermal equilibrium and triggers a thermal instability. The resulting condensations fall back towards the surface. They exhibit quasi-periodic behaviour, with periods longer than typical periods for solar coronal rain. We find line-of-sight velocities for these clumps in the range $50 \ \mathrm{km} \ \mathrm{s}^{-1}$ (blue shifted) to $250 \ \mathrm{km} \ \mathrm{s}^{-1}$ (red shifted). These are typical of those inferred from stellar H$\alpha$ line asymmetries, but the inferred clump masses of $3.6\times 10^{14}\ \mathrm{g}$ are significantly smaller.  We find that a maximum of $\simeq~3\%$ of the coronal mass is cool clumps. We conclude that coronal rain may be common in solar like stars, but may appear on much larger scales in rapid rotators.
\end{abstract}


\section{Introduction}

The interplay between heating and cooling is at the heart of the self-regulating cycle that determines the mass structure and dynamics of stellar coronae. Raising the heating drives material up from the large mass reservoir in the chromosphere. This increases the coronal density, pushing it out of thermal equilibrium. The resulting thermal instabilities trigger catastrophic cooling and the formation of condensations that rain out of the corona, lowering the overall coronal mass again. These heating and cooling cycles naturally act to regulate the amount of mass supported in the stellar corona \citep{2020PPCF...62a4016A,2022FrASS...920116A}.

Condensations are seen in two forms in the solar corona, both in the form of coronal rain and the much larger prominences that may be seeded by these condensations. On other solar-like stars, both of these types of coronal condensations are also observed. The most extensively studied are the large ``slingshot prominences'' observed on rapidly-rotating cool stars \citep{collier1989I,collier1989II}. These are clouds of cool ($T~\simeq~8000-10000 \ \mathrm{K}$) mainly neutral gas trapped in the hot ($T~\simeq~1-10 \ \mathrm{MK}$) stellar corona. They are detected as travelling absorption features that move through the line profiles of H$\alpha$ and many transition-region lines, and are seen in both single and binary stars \citep{collier1992,Hall1992,Byrne1996,Eibe1998,Barnes2000,Barnes2001,Petit2005,Dunstone2006,DunstoneThesis2008,Skelly2008,Skelly2009,2016MNRAS.463..965L,2020A&A...643A..39C,2021A&A...654A..42C,2021MNRAS.504.1969Z}. In a few cases, these features can be detected in emission, moving at velocities greater than the equatorial projected rotational velocity of the star. They often re-appear at the same rotation phase on the next stellar rotation, suggesting that they are co-rotating with the star. Their formation times and lifetimes are of order days. The drift rates of these features through the line profile provides their distance from the stellar rotation axis, showing that they form typically at (or beyond) the centrifugal co-rotation radius, where the outward centrifugal force dominates over the inward pull of the star's gravity. Masses derived from spectra (typically 10$^{14}$kg) are similar to those predicted from static, steady state models that use stellar surface magnetic maps as inputs \citep{2018MNRAS.475L..25V,2019MNRAS.483.1513W,2019MNRAS.485.1448V,2021MNRAS.505.5104W}. Their ejection may provide a significant contribution to the mass and angular momentum loss in the stellar wind \citep{2020MNRAS.491.4076J,2022MNRAS.513.5611F,2022MNRAS.514.5465W}. Surveys searching for the stellar equivalent of solar coronal mass ejections, however, have often struggled to detect them \citep{2014MNRAS.443..898L}.

More recently, it appears that the stellar equivalent of solar coronal rain has also been observed. H$\alpha$ asymmetries have been detected in surveys of many cool stars \citep{1993A&A...274..245H,1999A&A...341..527E,2018A&A...615A..14F,2019A&A...623A..49V,2022ApJ...926L...5N,2022NatAs...6..241N,2022ApJ...928..180W,2022ApJ...925..155K,2023ApJ...948....9I,2023ApJ...945...61N}. The lifetimes of these features are not known as they are not tracked over many stellar rotations, but their velocities are indicative of cooled gas travelling at speeds that are typically well above the rotational velocity of the corona, but below the escape speed, although they can in some cases exceed this value.  The fastest-moving condensations may have been detected while they are in the process of centrifugal ejection from the star. Recent analysis of such events on the M dwarf V374 Peg derived higher temperatures for these escaping features than have been derived for the trapped ``slingshot prominences'', suggesting that they may have been heated during their ejection \citep{2022MNRAS.513.6058L}. A range of masses is derived for these clouds, with typical values of $10^{13} - 10^{14}$ kg \citep{2019A&A...623A..49V}. While the relationship between these features and the relatively stable ``slingshot prominences'' is not clearly known, it is possible that when condensations form, they fall under gravity, either collecting in larger accumulations as prominences if they encounter a location where they can be supported, or falling freely towards the stellar surface as coronal rain.


Coronal rain has been extensively studied on the Sun since it was first observed \citep{kawaguchi1970, Leroy1972} and has been modelled at length in multi-dimensional simulations for a number of decades \citep{Antiochos1999, Karpen2006, Antolin2010, Froment2018, Li2022}. Recently, there has been extensive work on the relationship between heating mechanisms and the production of coronal rain in 2.5D and 3D simulations by \cite{Xia2012, Fang2013, Zhou2021, 2021ApJ...920L..15R}. This has allowed for the fine structure of coronal rain and the corona's response to its formation to be studied in unprecedented detail. The studies thus far demonstrate that coronal rain can occur in limit cycles known as Thermal Non-Equilibrium (TNE) and that the corona responds to coronal rain dynamics through rebound shocks and upflows from the chromosphere. It has also been demonstrated that the rain is embedded in an intricate environment complete with siphon and shear flows, while the rain blobs themselves have a steep temperature gradient reminiscent of the chromosphere-corona transition region \citep{Fang2015}.
 
Modelling these complex processes within stellar coronae is even more challenging, since it couples a large range of spatial and temporal scales. The solar literature is rich with simulations modelling the full 3D structure of the solar atmosphere \citep{Mikic1999,Lionello2009,Downs2010,Holst2014}, but in the case of other stars, however, there are fewer examples. To date, modelling of stellar condensations has focused on the larger, more stable ``slingshot prominences''  and on the contribution that their ejection (or their associated coronal mass ejections) might make to stellar spin down \citep{2013AN....334...77A,2017ApJ...840..114C}. Studies of stellar winds and rotational evolution often under-predict the torque that the wind alone can provide \citep{2012ApJ...754L..26M,2015ApJ...798..116R,2017MNRAS.466.1542S,2018MNRAS.474..536S,2018ApJ...854...78F} but models of  ``slingshot prominences'' suggest that their ejection may contribute to the loss of angular momentum \citep{2019MNRAS.485.1448V,2021MNRAS.505.5104W,2022MNRAS.514.5465W, 2022MNRAS.513.5611F}. 

The detection of numerous dynamic features on many stars suggests, however, that condensations may be present {\it throughout} the stellar corona and that they may play the same role as solar coronal rain in regulating the mass of the star's corona.  Their relationship to stellar flares is currently being explored through simultaneous observations of flares and H$\alpha$ line asymmetries \citep{2022ApJ...926L...5N,2022NatAs...6..241N,2022ApJ...928..180W,2022ApJ...925..155K,2023ApJ...948....9I,2023ApJ...945...61N}. In order to bridge the observational gap between solar and stellar studies, statistical \citep{2022ApJ...939...21M} and {\it Sun-as-a-star} observations have been conducted, showing red- and blue-shifted H$\alpha$ asymmetries associated with solar flares \citep{2022NatAs...6..241N,2022ApJ...939...98O}. These however did not reach the velocities that would allow them to escape from the solar corona.

The energy output from a flare does not simply contribute to heating the coronal gas. Studies of solar flares suggest that the majority contributes to mass ejection. There is a clear correlation between the kinetic energy of ejected mass and the energy that results from heating of the gas, typically seen in X-ray emission. Extending this correlation to the typical X-ray energies seen in stellar flares, however, predicts stellar mass ejections with an unphysically-large kinetic energy \citep{2013ApJ...764..170D}. In order to explain this apparent breakdown of the solar relationship, \cite{2018ApJ...862...93A} suggested that the strong magnetic fields of rapidly-rotating stars with powerful flares may inhibit mass ejection. Detections of stellar prominences (and coronal mass ejections) appear to support the hypothesis that stellar mass ejections carry away less kinetic energy than their solar counterparts, relative to the overall flare energy \citep{2014MNRAS.443..898L,2016MNRAS.463..965L,2017IAUS..328..198K,2020MNRAS.493.4570L,2019ApJ...877..105M}. 

In this paper we present the first part of a larger-scale study of the formation and dynamics of condensations within stellar coronae. We identify two different types of condensations: the quasi-stable accumulation of material, such as centrifugally-supported ``slingshot prominences'' and the more dynamic condensations that fall under gravity as coronal rain. These two behaviours have similarities to the ``centrifugal'', or ``dynamical'' magnetospheres of massive stars \citep{Sundqvist2012, Petit2012, Villarreal2017}. In this initial study we focus on the smaller-scale coronal rain that may form throughout the corona, Studies of the larger-scale ``slingshot prominences'' will be presented in later papers. We select the stars most likely to be magnetically active, with many flares. These are the most X-ray luminous, rapidly-rotating stars. In this initial study, we demonstrate the formation of these condensations in response to excess heating and analyse their subsequent evolution.


\section{Modelling}

\subsection{Magnetohydrodynamics}

We solve the equations of magnetohydrodynamics (MHD) in the star's rotating frame. These equations are:

\begin{equation}
	\partial_{t} \rho + \nabla \cdot (\rho \pmb{v}) = 0
	\label{eq:mass_cons}
\end{equation}

\begin{equation}
	\partial_{t} (\rho \pmb{v}) + \nabla \cdot (\rho \pmb{v} \pmb{v} - \pmb{B} \pmb{B} + p_{\rm{tot}}\rm{\pmb{I}}) = \rho \pmb{\textsl{g}}_{\rm{eff}}
	\label{eq:mom_cons}
\end{equation}

\begin{equation}
	\partial_{t} \pmb{B} + \nabla \cdot (\pmb{v}\pmb{B} - \pmb{B}\pmb{v}) = 0
	\label{eq:mag_cons}
\end{equation}

\begin{equation}
	\partial_{t} e + \nabla \cdot ((e + p_{\rm{tot}}) \pmb{v} - \pmb{B} \pmb{B}  \cdot \pmb{v}) = \rho \pmb{\textsl{g}}_{\rm{eff}} \cdot \boldsymbol{v} + \nabla \cdot \boldsymbol{F}_{\rm{c}} + Q,
	\label{eq:en_cons}
\end{equation}
here $\rho$, $\bm{v}$, $\bm{B}$, $p_{\rm{tot}}$, $e$, $\boldsymbol{F}_{\rm{c}}$, $\bm{g}_{\rm{eff}}$ and $Q$ are, respectively, the density, velocity, magnetic field, total gas pressure, internal energy, conductive heat flux, effective acceleration due to gravity, rotation and finally the difference between phenomenological heating and optically thin radiative cooling.

Here the effect of rotation has been combined with gravitational acceleration to give an effective gravity: 
\begin{equation}
	\pmb{\textsl{g}}_{\rm{eff}} = \frac{-G M_{\ast}}{r^{2}} + \Omega_{\mathrm{fr}}^{2} r \sin^2(\theta) \hat{\boldsymbol{r}} +
    \Omega_{\mathrm{fr}}^{2} r \sin(\theta) \cos(\theta) \hat{\boldsymbol{\theta}}
\end{equation}
where $\Omega_{\mathrm{fr}}$ is the angular velocity of the rotating frame, which in this case is the stellar rotational velocity. The distance from the origin is $r$ and $\theta$ is the poloidal angle. We define the rotational axis to be along the $z$-direction, following the convention used in spherical geometry. The thermal conductive flux is defined as:
\begin{equation}
    \vec{F}_{\rm{c}} = - \kappa_{0} T^{5/2} \hat{b} \hat{b} \cdot \vec{\nabla} T,
	\label{eq:thermal_cond}
\end{equation}
where $\kappa_{0} = 8 \times 10^{-7}$ cgs is the Spitzer thermal conduction coefficient. To deal with the challenge of resolving the steep transition region temperature gradients, we employ the Transition Region Adaptive Conduction (TRAC) method as implemented in MPI-AMRVAC. See the works of \cite{Johnston2019, Johnston2020, Iijima2021} for details. This means applying a modified thermal conduction coefficient to broaden the transition region for temperatures below a specified cutoff temperature. The contributions from heating and cooling are:
\begin{equation}
	Q = Q_{\rm{h}} + Q_{\mathrm{loc}} - n_{\rm{e}} n_{\rm{i}} \Lambda(T),
	\label{eq:cooling}
\end{equation}
where $Q_{\rm{h}}$ is the steady background heating and $n_{\rm{e}}$ and $n_{\rm{i}}$ are the electron and ion number densities, respectively. $Q_{\mathrm{loc}}$ is an impulsive localised heating source and $\Lambda(T)$ is the optically thick radiative loss function for which we use the \textit{SPEX\_DM} curve \citep{Schure2009, Hermans2021}. The exact form of the heating sources and the role they play is explained in the following section.

\subsection{Wind modelling}

\begin{table*}
	\centering
	\caption[]{Stellar and simulation parameters.
	\label{tab:parameters}}
	\begin{tabular}{cccc}
		\hline
		Name & Parameter & Value & Source \\
		\hline
 		Radius & $R_{\ast}$ & 1 $R_{\sun}$ & - \\
		Mass & $M_{\ast}$ & 1 $R_{\sun}$ & - \\
		Period & $P_{\ast}$ & 1 day & - \\
        Polar magnetic field strength & $B_{0}$ & 100 G & - \\
        Chromosphere height & $h_{\rm{ch}}$ & 5 Mm & - \\
		Background heating length-scale & $\lambda_{0}$ & 40 Mm & \cite{Downs2010} \\
		  Wind heating length-scale & $\lambda_{\rm{w}}$ & 0.7 $R_{\sun}$ & \cite{Downs2010} \\
		  Background heating amplitude & $H_{0}$ & $7.28 \times 10^{-5} \ \mathrm{erg} \ \mathrm{cm}^{-3} \ \mathrm{s}^{-1}$ & \cite{Downs2010} \\
		  Wind heating amplitude & $H_{\rm{w}}$ & $5 \times 10^{-7} \ \mathrm{erg} \ \mathrm{cm}^{-3} \ \mathrm{s}^{-1}$ & \cite{Downs2010} \\
        Local heating length-scale& $\lambda_{\rm{h}}$ & 50 Mm & \cite{Zhou2021} \\
		  Local heating amplitude & $H_{\mathrm{loc}}$ & $2 \times 10^{-2} \ \mathrm{erg} \ \mathrm{cm}^{-3} \ \mathrm{s}^{-1}$ & \cite{Zhou2021} \\
        Local heating height & $r_{\mathrm{h}}$ & 6.5 Mm & \cite{Zhou2021} \\
        Relaxation time & $t_{\mathrm{relax}}$ & 1000 s & \cite{Zhou2021} \\
        Ramp time & $t_{\mathrm{ramp}}$ & 500 s & \cite{Zhou2021} \\
        Local heating separation & $\theta_{\rm{loc}}$ & $\pi/2$ radians & - \\
		Local heating width & $\sigma$ & 1 $R_{\sun}$ & - \\
		\hline
	\end{tabular}
\end{table*}

\subsubsection{The extended wind}

We establish a corona and drive an extended wind by applying a scale height heating model. This takes an exponential form and consists of two terms, one to heat the chromosphere and lower corona and a second maintaining the extended wind. This takes the form:
\begin{equation}
	Q_{\rm{h}} = H_{0} \exp \left( -\frac{r - R_{\ast}}{\lambda_{0}} \right) + H_{\rm{w}} \exp \left( -\frac{r - R_{\ast}}{\lambda_{\rm{w}}} \right),
	\label{eq:heating}
\end{equation}
where $H_{0}$ and $\lambda_{0}$ are the amplitude and length scale of the heating responsible for the lower corona and $H_{\mathrm{w}}$ and $\lambda_{\mathrm{w}}$ are the amplitude and length scale of the heating that maintains the extended wind. This form of heating is used by \cite{Lionello2009} and \cite{Downs2010}.	

\subsubsection{Local heating}
\label{sec:local_heating}

To accompany the background heating we apply localised footpoint heating. This is responsible for forcing the system into TNE. The form of heating we use is adapted from \cite{Zhou2021}. The only difference is that we apply it in spherical geometry, with the result that the heating separation is expressed in radians.

The heating takes the form of two Gaussian profiles separated by an angle $\theta_{\rm{loc}}$ such that $\theta_{\rm{r}} = (\pi + \theta_{\rm{loc}})/2$ and $\theta_{\rm{\ell}} = (\pi - \theta_{\rm{loc}})/2$ are the the centres of the two Gaussian profiles. This ensures that the footpoints of the flaring loop are symmetric about the star's magnetic equator. Both footpoint profiles have a Gaussian width of $\sigma$:
\begin{multline*}
	Q_{\rm{loc}} = H_{\mathrm{loc}} R(t) C(r) \\ \left[ \exp \left( \frac{-(\theta - \theta_{\rm{r}})^{2}}{\sigma^{2}} \right) + \exp \left( \frac{-(\theta - \theta_{\rm{\ell}})^{2}}{\sigma^{2}} \right)  \right].
\end{multline*}
The two coefficients, $R(t)$ and $C(r)$, moderate the heating input by applying a relaxation time, $t_{\rm{relax}}$, and a ramp time, $t_{\rm{ramp}}$. This allows the system to relax from the initial conditions and slowly increases the heating in a manner that does not produce unphysical flows. These conditions are summarized as:
\begin{equation}
    R(t) = 
\begin{cases}
    0, & \text{if } t \le t_{\rm{relax}}\\
    (t-t_{\rm{relax}})/t_{\rm{ramp}}, & \text{if } t_{\rm{relax}} < t < t_{\rm{relax}} + t_{\rm{ramp}} \\
	1, & \text{if } t \ge t_{\rm{relax}} + t_{\rm{ramp}}.
\end{cases}
\end{equation}
The footpoint heating also has a Gaussian profile in the radial direction of the form:
\begin{equation}
    C(r) = 
\begin{cases}
    1, & \text{if } r \le r_{\mathrm{h}} \\
    \exp \left( -(r - R_{\ast})^{2}/\lambda_{\mathrm{h}}^{2} \right), & \text{if } r > r_{\mathrm{h}}.
\end{cases}
\end{equation}

As we are exploring are large scale coronal rain, the choice of heating separation needs to result in a thermalised loop that maximises the furthest extent of a condensation. For a dipole magnetic field geometry, this requirement would result in maximising the separation angle. However, this is also constrained by the requirement for the thermalised loop to be located within the closed field region. Since the stellar wind opens up the field close to the poles, this limits the separation angle and for this simulation we keep $\theta_{\mathrm{loc}} = \pi/2$. Once the heating is established and $t_{\rm{ramp}}$ has passed, the local heating is constant and persists for the remainder of the simulation. This is consistent with solar magnetic arcade simulations in the literature \citep{Xia2012, Fang2013, Zhou2021}. We note that the heating described above fulfills the necessary conditions for TNE, that the heating needs to be low in the corona and quasi-steady \citep{Klimchuk2019a, Klimchuk2019b}.


All the variables used for the heating modelling and to define the physical parameters of our simulated star are summarized in Table.~\ref{tab:parameters}.

\section{Numerical modelling}

We solve the MHD equations using the parallel, block based, adaptive mesh code MPI-AMRVAC \citep{Xia2018,Keppens2021}. Time and space discretisation are handled with the strong stability preserving third-order Runga-Kutta timestepping and the second order van Leer limiter \citep{vanLeer1974} respectively. We pair these methods with the HLL Riemann solver \citep{Harten1983} to complete the advection algorithm. Additionally we use the method of super time stepping to address thermal conduction and apply the background field splitting strategy to improve code stability. Finally, the divergence of the magnetic field is handled according to Powell's \textit{8-wave} formalism \citep{Powell1999}.

\subsection{Computational grid}

Our computational grid extends between $r~\in~\{1, 50\} \ R_{\ast}$ and $\theta~\in~\{0, \pi\} \ \mathrm{radians}$. This places the outer radial boundary sufficiently far from the stellar surface to avoid impacting the dynamics in the lower corona and  facilitates the establishment of a stable stellar wind.

A base resolution of 128 cells in the radial direction and 96 cells in the poloidal direction, with the addition of 4 refinement levels, is used to provide an effective resolution of $2048~\times~1536$. Grid cells in the radial direction are stretched, allowing us to concentrate resources in the chromosphere and transition region, while not needlessly over-resolving the smooth regions of the extended wind.

\subsubsection{Initial conditions}

We initialise the grid with a hybrid approach. The transition region is treated as a discontinuity separating a hydrostatic chromosphere extending out to $r~=~5 \ \rm{Mm}$ above the stellar surface. Beyond this point, and out to the upper boundary at $r~=~50 \ R_{\ast}$, a Parker wind solution is used. We keep density, pressure and velocity constant in the poloidal direction.

This initial condition is designed to provide a smooth density profile paired with the discontinuous jump in temperature characteristic of the transition region. Either side of the transition region there is an isothermal profile, such that $T~=~1 \times 10^{4} \ \rm{K}$ in the chromosphere and $T~=~1.7 \times 10^{6} \ \rm{K}$ in the Parker wind region. One issue this setup presents is a nonphysical jump in pressure at the transition region. This is, however, dissipated early in the simulation and has no impact on the solution at the time when the flow begins forming condensations. 

Our simulated chromosphere has an initial width $h_{\mathrm{ch}}~=~5~\mathrm{Mm}$ and we specify the number density at the top as $10^{9} \ \mathrm{cm}^{-3}$ 
and the constant, isothermal temperature as $10^{4} \ \mathrm{K}$. The pressure and density profiles are determined by integrating the hydrostatic equation, from the top of the chromosphere down to the surface of the star. Above $h_{\mathrm{ch}}$, the Parker wind equation is solved for pressure and density, leading to the pressure jump described above.  The precise values of the initial density and temperature of the upper chromosphere do not impact the simulation simulation once it reaches steady state. This is because the position, density and temperature of the chromosphere-coronal transition self-consistently adjust based on the heating rate (c.f. equation \ref{eq:cooling}) and local conditions. Therefore, the values in the simulation are correct for young active stars insofar as the heating prescription is correct.

The magnetic field is initialised as a dipole aligned with the rotational axis of the star. For the current simulation, we kept the dipole field strength constant at $100 \ \mathrm{G}$ (polar value).  This value is based on the empirical trends reported by \cite{Vidotto2014} for low-mass stars and represents a typical value for a star of this class with a one day rotation period.

In summary, our simulated star differs from the inactive, present day Sun, in that it has a higher rotation rate and a consequently higher magnetic field strength.

\subsubsection{Boundary conditions}

We follow the approach of \cite{Zhou2021} when specifying our chromospheric boundary. This means extrapolating the hydrostatic equations into the boundary such that pressure and density are constant. The velocity is set to zero and the magnetic field is fixed to the initial configuration. At the maximal radial boundary, all quantities are outflowing. The poloidal boundaries are set to be axisymmetric. This means that all quantities are set to be reflective, except for the non-axial components of the vector fields, which are set to be asymmetric (i.e. with the same value but with negative sign). This ensures zero flux through the poloidal boundaries. 



\section{Results}

\begin{figure*}
	\centering
	\includegraphics[width=0.49\textwidth,trim={0cm, 0cm, 0cm, 0cm},clip]{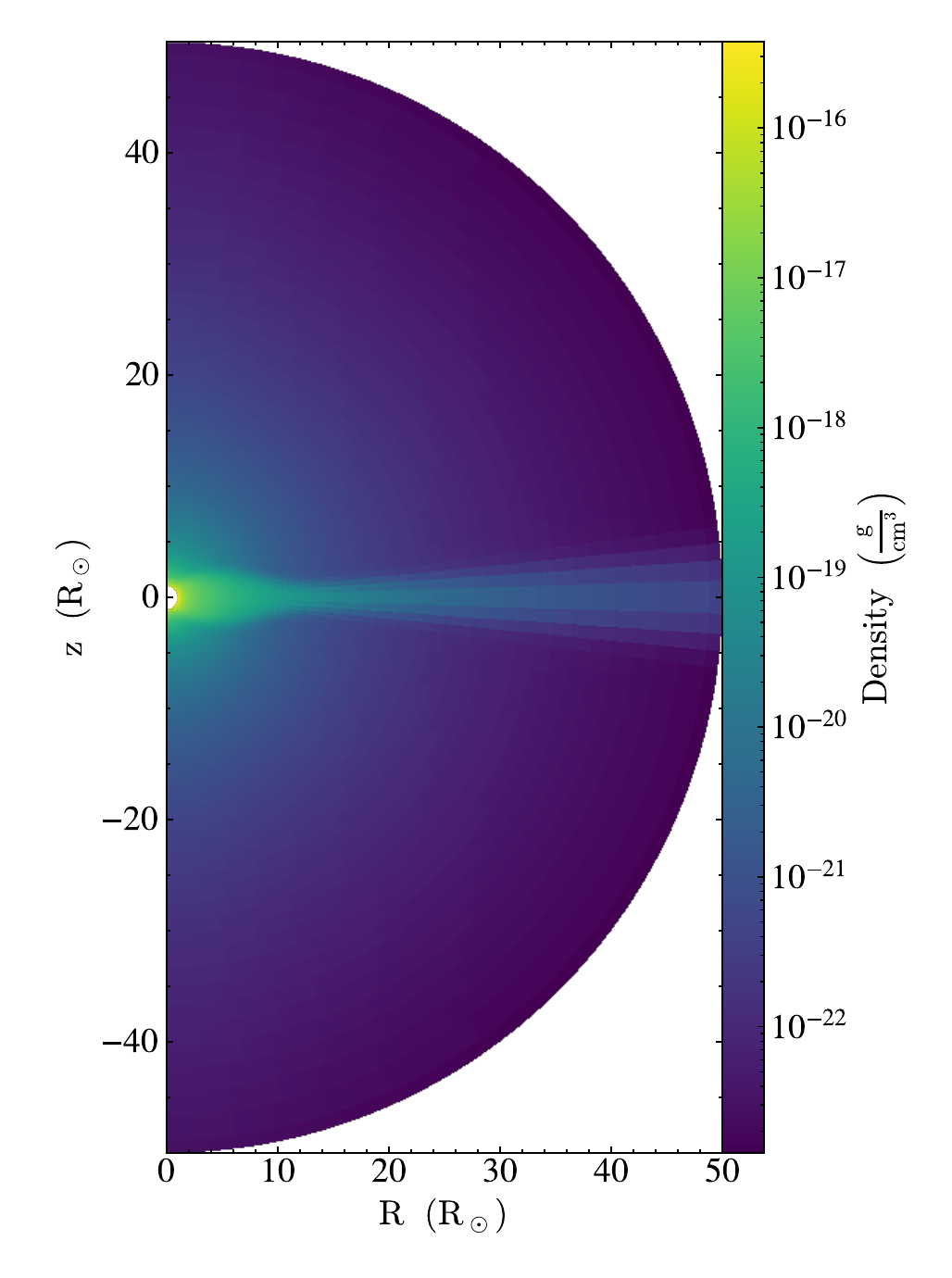}
	\includegraphics[width=0.49\textwidth,trim={0cm, 0cm, 0cm, 0cm},clip]{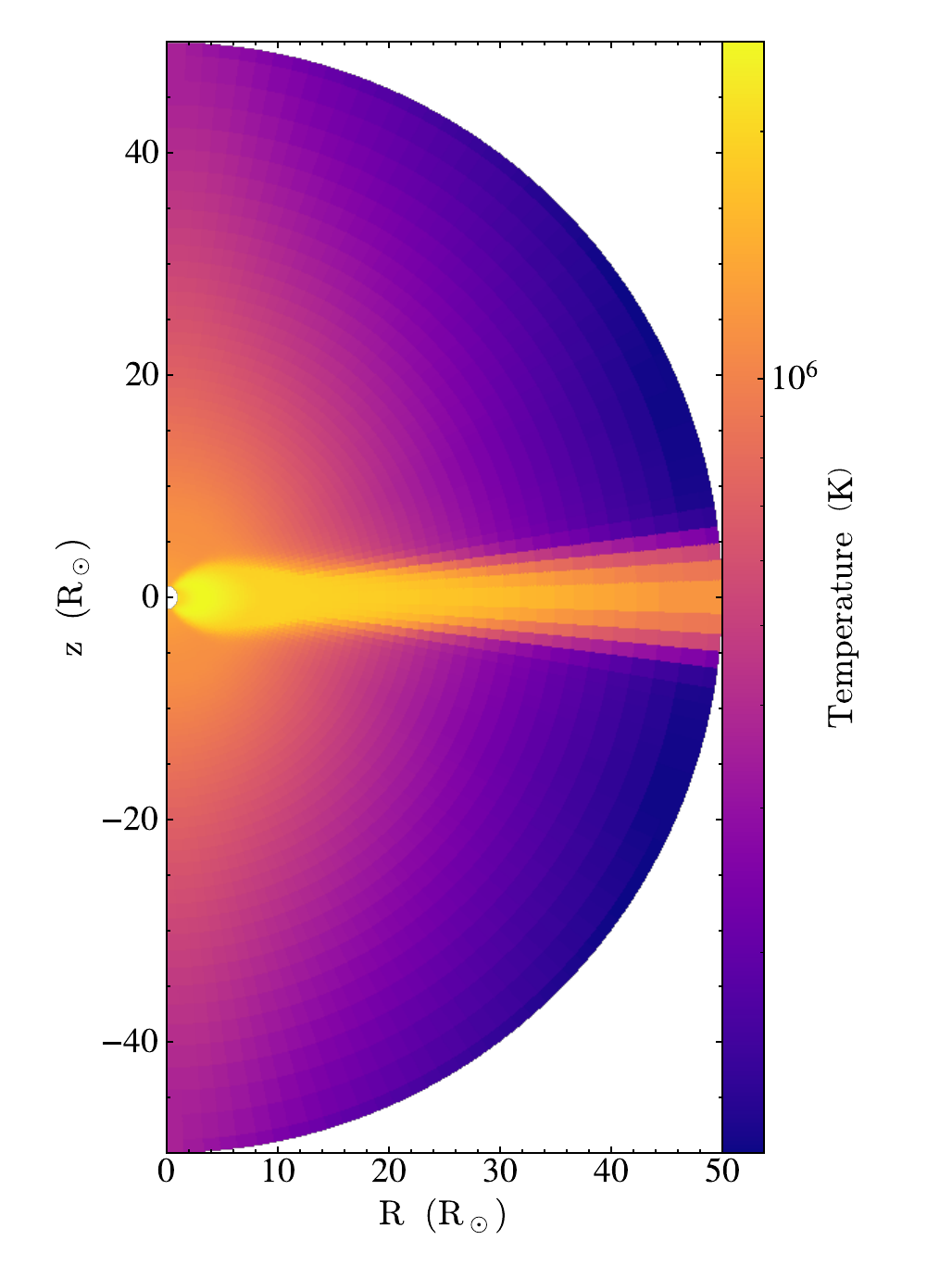}
	\caption[]{Density and temperature steady state profiles for the entire grid, obtained by advancing the simulation's initial conditions with only background heating until all quantities no longer change. \label{fig:init_conds}}
\end{figure*}

\begin{figure*}
	\centering
	\includegraphics[width=0.99\textwidth,trim={0cm, 0cm, 0cm, 0cm},clip]{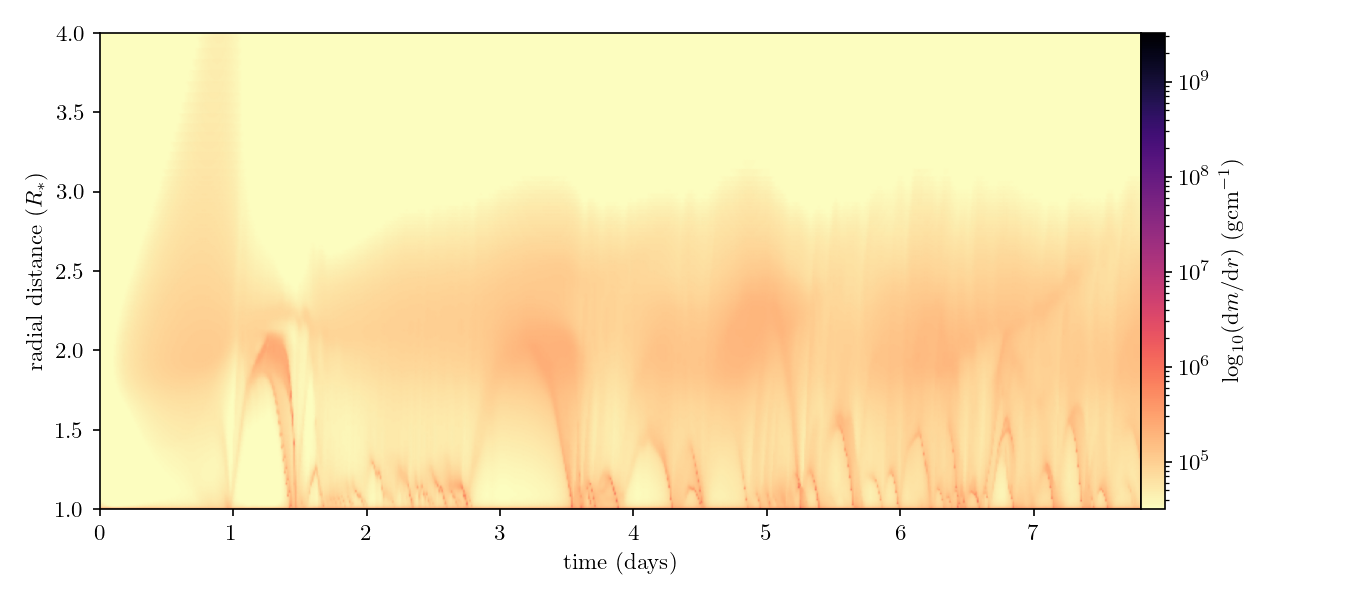}
	\caption[]{Radial mass distribution for the inner 4 $R_{\ast}$ of the simulation. The impact of the initial conditions is apparent on the left, where mass can be seen moving to larger radii as time increases. This mass is removed by the stellar outflow within $\simeq2$ days. Note the distinct arch of material tracking up and back to the stellar surface during this time. This represents the first condensation forming. From this point, multiple arching tracks of varying heights above the surface can be seen. Each of these marks a condensation event and a subsequent fall to the stellar surface where it is reabsorbed by the chromosphere. Large arches, reaching up to $\simeq2.5 \ R_{\ast}$, represent condensations forming and tracking through the full thermalised loop extent. Smaller arches indicate the presence of a condensation that forms low in the legs of the loop. The maximum arch height effectively marks the source surface for the stellar wind. Beyond this radius the plasma is a free streaming, radially outflowing wind.\label{fig:radial_mass}}
\end{figure*}

In the following sections we lay out the results, starting with an overview of mass density and temperature for both the simulation initial conditions and for a significant condensation event, happening at $\sim~6.7$ days into the simulation. We follow with a time series analysis of the radial distribution of mass within the magnetosphere's source surface and end by characterising the mass and velocity properties of the aforementioned condensation event, placing it into an observational context.

\subsection{Global magnetosphere structure}

The initial conditions for the magnetosphere and extended wind are shown in Fig. \ref{fig:init_conds}. There is clear division between open, free-streaming wind and slow, higher-density wind, concentrated towards the magnetic equator. This structure is consistent with the solar wind.

Close to the star, the so called \textit{dead zone} of mostly hydrostatic gas can be seen extending out to $\sim~10 \ R_{\ast}$. This \textit{dead zone} represents the region of closed field lines. In the limit of ideal magnetohydrodynamics, there should be zero cross-field velocity. However, there is non-zero numerical diffusion due to relatively course resolution at the boundary between the open and closed field. This diffusion acts in the manner of resistivity and leads to reconnection of the field and material outflow from the \textit{dead zone}, into the extended wind, forming the star's \textit{slow wind}. This global structure is consistent with early numerical wind models of the solar wind \citep{Wang1998, Keppens1999, Lionello2001} and with wind models of cools stars \citep{Reville2015, Pantolmos2017, Finley2017} more generally. In our case the \textit{dead zone} extends further from the stellar surface than in simulations of the solar wind. This is expected, however, as our magnetic field strength is an order of magnitude larger than that typically used in solar wind simulations.

\subsection{Local heating and the transition to thermal non-equilibrium}

\begin{figure*}
	\centering	
	\includegraphics[width=0.99\textwidth,trim={0cm, 0cm, 0cm, 0cm},clip]{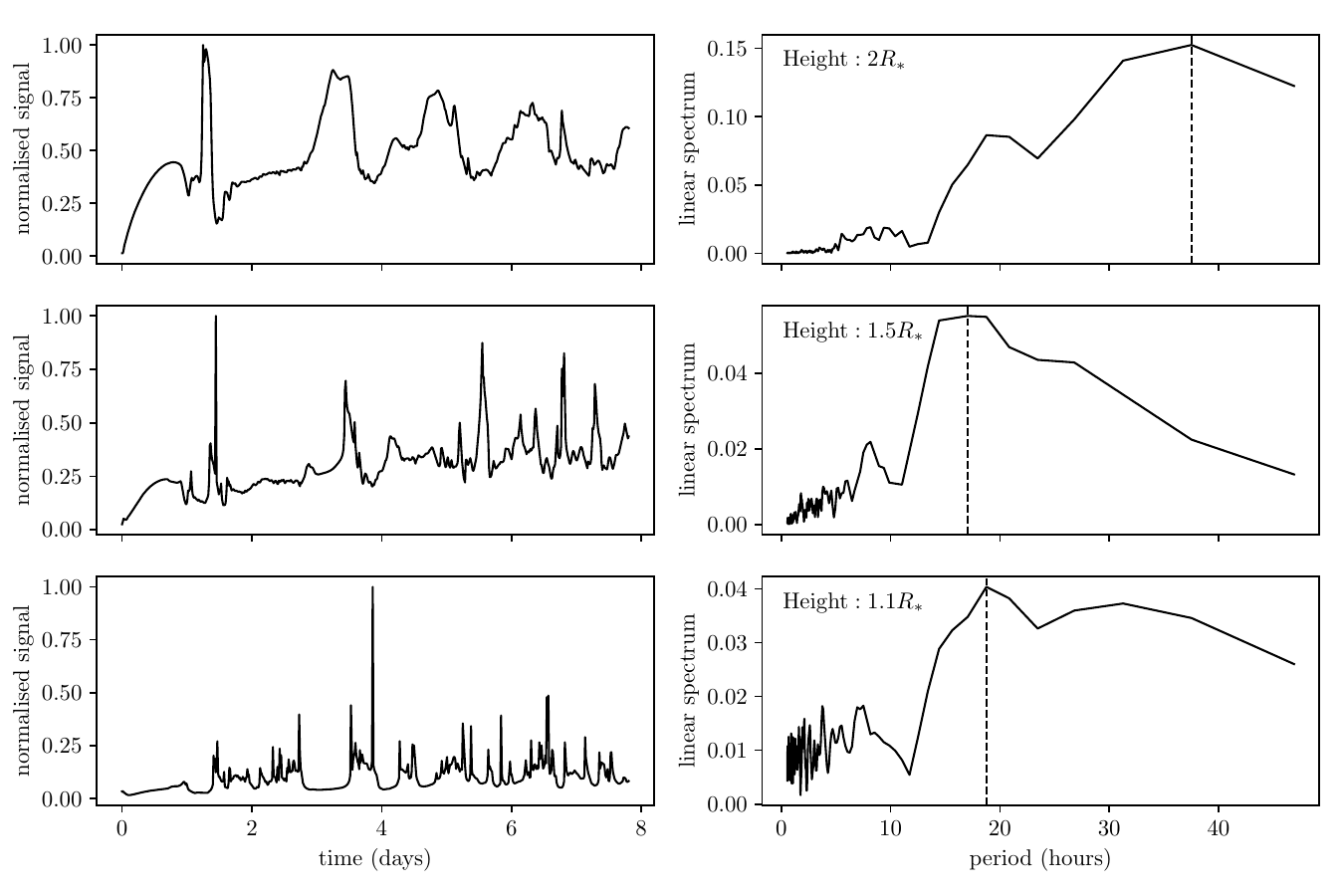}
	\caption[]{The left panel shows the mass variation with time at three separate heights: $2.0 \ R_{\ast}, 1.5 \ R_{\ast}$ and $1.1 \ R_{\ast}$, while the right panel shows the corresponding power spectrum of these variations. The period of the peak power P$_{\rm Peak}$ corresponding to each radial slice and its height above the surface are, from top to bottom: P$_{\rm peak}$ ($2 \ R_{\ast})~=~37 \ \mathrm{hr}$, P$_{\rm peak}$($1.5 \ R_{\ast})~=~17 \ \mathrm{hr}$, P$_{\rm peak}$($1.1 \ R_{\ast}) ~=~18.7 \ \mathrm{hr}$. Each P$_{\rm peak}$ is indicated by a vertical dashed line. \label{fig:spectrum}}
\end{figure*}

Once the simulation has started we allow the steady state profile to persist for a short time, $t_{\rm{relax}}$, before commencing the localised heating. The amplitude of $Q_{\mathrm{loc}}$ is then linearly increased to full strength over a time $t_{\rm{ramp}}$. This allows the system to adjust to the new heating profile without causing significant numerical problems.

After this initial phase, the density inside the thermalised loop begins to increase, filling with evaporated chromospheric material. Radiative losses therefore increase and the gas begins to cool, leading to the local onset of thermal instability, and the gas condenses into coronal rain. This condensed mass then drains from the thermalised loop by descending to one of the footpoints and is finally absorbed by the chromosphere. This process is cyclic and these stages correspond to the three steps of the TNE cycle as described by \cite{Antolin2019}. 

The local heating method detailed in Section \ref{sec:local_heating} leads to the establishment of a thermalised loop extending into the closed magnetosphere. After $\sim~1$ day of simulated time the first condensation event occurs. This is highly symmetric and largely a reflection of the initial conditions. After this initial event is reabsorbed into the chromosphere ($>~1.5$ days), symmetry is broken and draining of the thermalised loop is asymmetric from here on.

The time evolution of these initial features and the subsequent condensations, their heights above the surface and time of re-absorption by the chromosphere, are characterised in Fig. \ref{fig:radial_mass}. This plot employs the method of \cite{ud-Doula2008} in which the mass density is integrated over the two angular directions ($\theta$ and $\phi$) leaving the mass distributed in the radial direction $\mathrm{d}m/\mathrm{d}r$. One can think of this radial mass distribution as the total mass contained in each spherical shell and at each value of $r$. Calculating $\mathrm{d}m/\mathrm{d}r$ for each time snapshot gives us a time series of the mass as a function of height above the stellar surface.

Visual inspection of Fig. \ref{fig:radial_mass} shows a steep drop off in $\mathrm{d}m/\mathrm{d}r$ at $\simeq~2.5 \ R_{\ast}$ (corresponding to $\simeq~1.5 \ R_{\ast}$ above the stellar surface). We interpret this as the {\it source surface}, the boundary between the closed and open magnetospheres. This value is similar to the solar value - a result that is surprising, given the relative strength of our dipole field compared to the Sun. We point out, however, that the rotation rate of our star is significantly higher than the Sun (1 day rather than $\sim~24$ days). This additional rotational acceleration helps to open up the magnetosphere, allowing material to escape to the free streaming wind that would usually be confined to the \textit{dead zone}.

Within the source surface, there are multiple arches of material accruing at different heights above the stellar surface. There are two families of such features, either large arcs that extent to $\simeq~2 \ R_{\ast}$ (the first symmetric condensation is the most apparent), and smaller arcs that represent material condensing in the legs of the thermalised loop. In this latter case, the arc is formed of material that is within a single loop leg. For larger arcs, condensing material can either fall within the same leg or traverse the entire loop. One characteristic of this simulation is that gas begins condensing low down in the loop while still traveling towards the apex. This is a departure from smaller scale or simplified 1D models, where condensations typically occur at or close to the loop apex. We note that this behaviour is seen when symmetric heating methods are used. Random or asymmetric heating can lead to condensations significantly away from the apex \citep{Li2022}.

We postulate that this difference is due to the extended length of our simulated loop compared to solar loops. The cooling time for our loop may be similar to a solar loop, but the typical travel time may be much greater due to the loop length. One would expect that the site along the loop where condensation occurs should correspond to the distance traveled in the cooling time. Our thermalised loop extends to $\simeq 1 \ R_{\ast}$ above the stellar surface. This gives a total loop length of $\simeq 2.76 \ R_{\ast}$ and therefore the half loop length is $\simeq 1.38 \ R_{\ast} = 960 \mathrm{Mm}$. This length is roughly an order of magnitude larger than solar coronal arcade simulations in the literature \citep{Fang2013, Xia2012, Zhou2021, Li2022}. 

\subsection{Characteristic timescales}

To place these results in context, we can estimate the relevant timescales for this system. Following \citet{2014LRSP...11....4R} we scale quantities to typical values as follows: $n_{9}$ is the typical loop number density in units of $10^{9} \ \mathrm{cm}^{-3}$, $L_{9}$ is the loop half length in units of $10^{9} \mathrm{cm}$ and $T_{6}$ is the typical loop temperature in units of MK. Typical numbers from our simulation would be $n_{9}~=~9.96$, $T_{6}~=~6.0$, and $L_{9}~=~96.06$ (for the tallest features in Fig. \ref{fig:radial_mass} that extend to $\simeq~ 1.0 \ R_{\ast}$ above the surface) or $L_{9}~=~9.6$ (for the lower features in Fig. \ref{fig:radial_mass} that extend only to $\simeq~0.1 - 0.2 \ R_{\ast}$).

With these value we find the free-fall time for a height of 1 $R_{\ast}$ above the stellar surface is $t_{ff}~\approx~1.1 \ \mathrm{hr}$. This is similar to the isothermal sound travel time 
\begin{equation}
    t_{\rm s} \approx  80 \frac{L_9}{T_6^{1/2}}
\end{equation}
which gives $t_{\rm s} \approx 0.9 \ \mathrm{hr}$ for this lengthscale. This suggests that the timescale on which a conduction front could reach the summit of the taller loops is similar to the timescale on which material could fall back to the surface.

This timescale is also similar to the radiative cooling time
\begin{equation}
    t_{\rm r} \approx  3000 \frac{T_6^{3/2}}{n_9}
\end{equation}
which gives $t_{\rm r} \approx 1.2 \ \mathrm{hr}$. By comparison, the timescale for conductive cooling (which depends on the lengthscale) is
\begin{equation}
    t_{\rm c} \approx  1500 \frac{n_9 L_9^2}{T_6^{1/2}}
\end{equation}
which gives $t_{\rm c} \approx 4.34 \ \mathrm{hr}$ for the shorter loop but the much longer value of  $t_{\rm c} \approx 434 \ \mathrm{hr}$ for the longer loop. Thus, for the longer loop, the dynamical and radiative cooling times are similar, although the conductive timescale is longer. This is consistent with our observation that for longer loops, some condensations may form within an upflow and fall back to the surface on similar timescales. These simple estimates suggest that the dynamics and energetics of these loops are closely intertwined. The length of the conductive timescale may point to a much longer recurrence time between condensation events than is seen on the Sun.

\subsection{Periodicities in condensation}

To further investigate the time period and cyclic behaviour of the condensation events and to determine any dominant periods in the rise and fall of condensing material in the thermalised loop, we analyse several slices at different heights through the radial mass distribution in Fig. \ref{fig:radial_mass}. The results are shown in Fig. \ref{fig:spectrum}. The heights of each investigated slice and the corresponding period of peak power $P_{\rm peak}$ are: $P_{\rm peak} (1.1 \ R_{\ast}) ~=~18.7 \ \mathrm{hr}$, $P_{\rm peak} (1.5 \ R_{\ast})~=~17 \ \mathrm{hr}$, $P_{\rm peak} (2 \ R_{\ast})~=~37 \ \mathrm{hr}$. Slices closer to the surface were chosen to coincide with the smaller condensation arcs and the higher slice to coincide with the larger arcs in Fig. \ref{fig:radial_mass}.

In the case of the upper slice at $2 \ R_{\ast}$, we find the dominant period is $37 \ \mathrm{hr}$. This appears to be the recurrence time between the large condensation events. In the case of the two lower slices at $1.1 \ R_{\ast}$ and $1.5 \ R_{\ast}$, we find the dominant periods are approximately half that of the period at $2 \ R_{\ast}$. This may either represent the lifetime of the largest condensation events, or simply the period doubling effect that results from both legs of the loop being counted as separate signals. We note that there are also many lower-amplitude signatures of much shorter periods on timescales of a few hours.
The dynamics of the condensed material in our simulation therefore appear to behave as a scaled-up solar loop, with the associated longer time scale for condensations ($37 \ \mathrm{hr}$), but also the shorter-period signature of coronal rain, similar to that in the Sun \citep{2022ApJ...931L..27S}. 

In the following section, we analyse one condensation event, looking at the mass and velocity behaviour and quantify how it fits into recent observations.

\subsection{Large condensation event}

\begin{figure*}
	\centering
	\includegraphics[width=0.49\textwidth,trim={0cm, 0cm, 0cm, 0cm},clip]{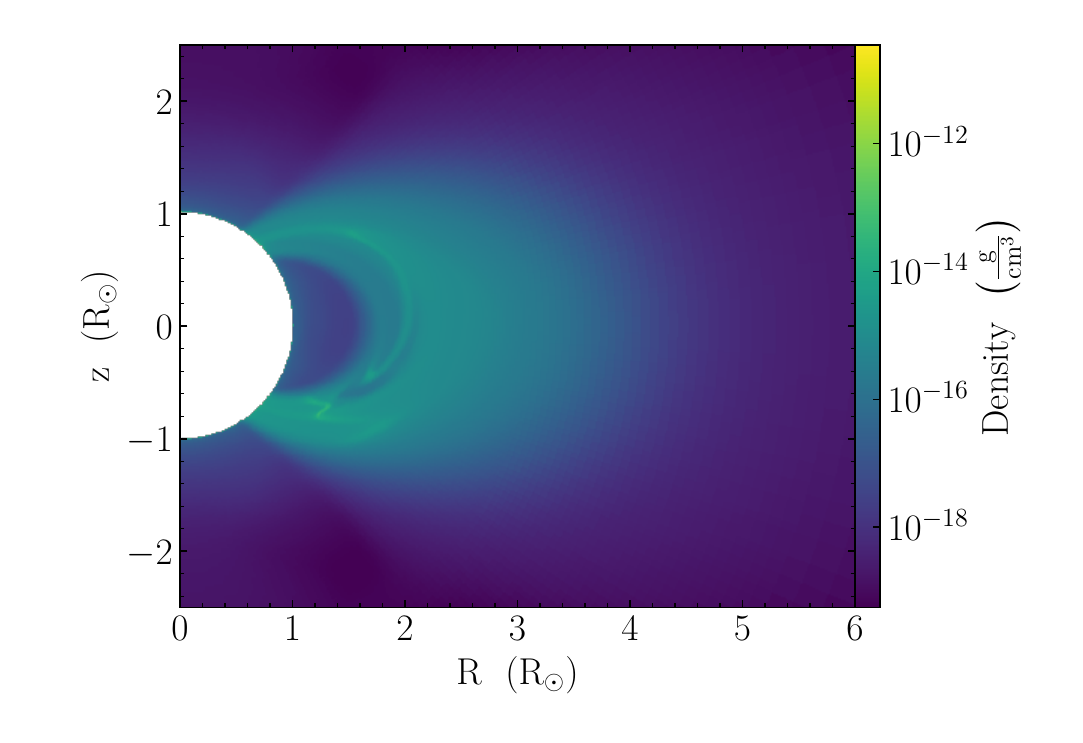}
	\includegraphics[width=0.49\textwidth,trim={0cm, 0cm, 0cm, 0cm},clip]{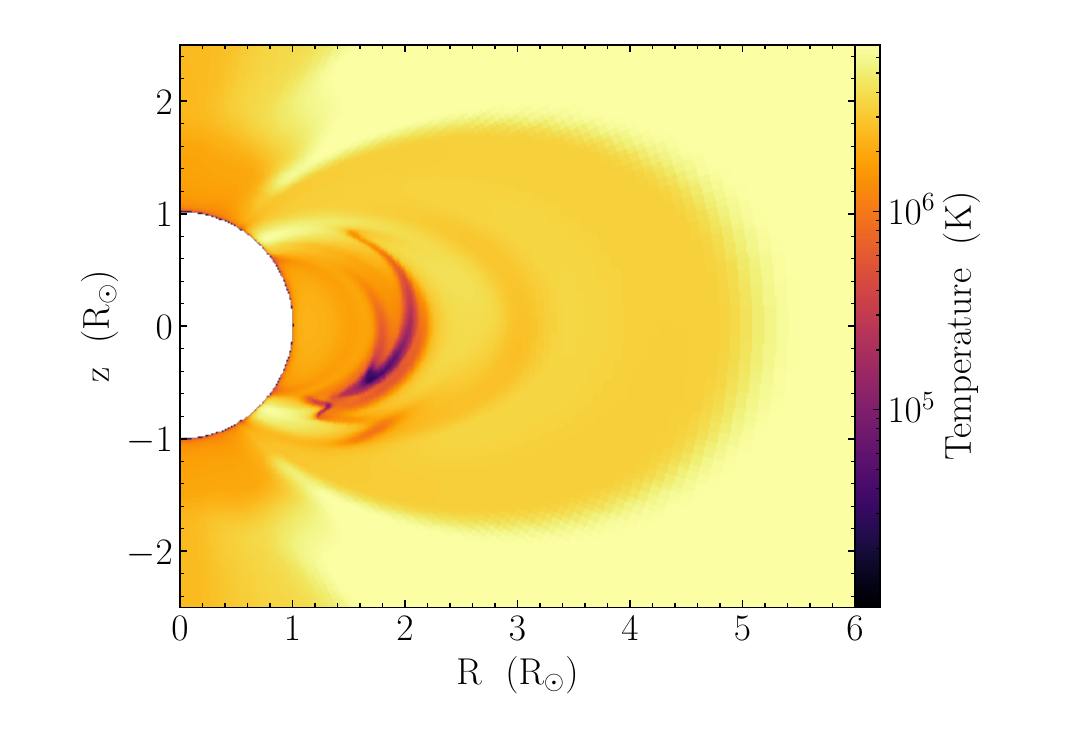}
	\includegraphics[width=0.49\textwidth,trim={0cm, 0cm, 0cm, 0cm},clip]{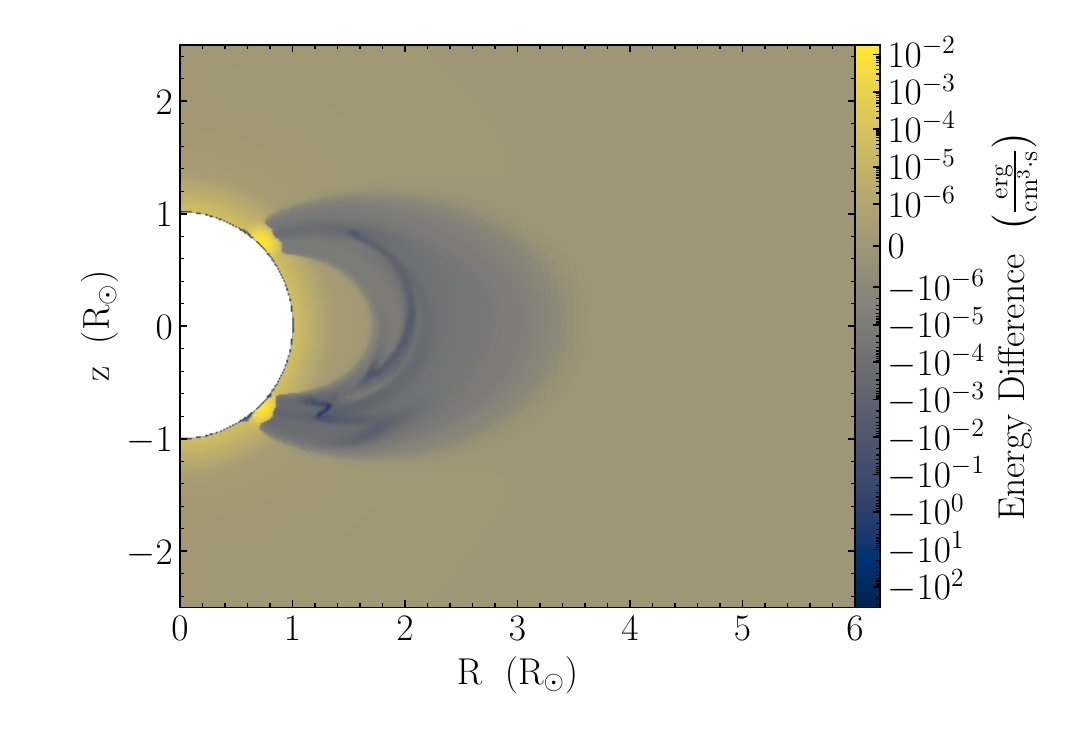}
	\includegraphics[width=0.49\textwidth,trim={0cm, 0cm, 0cm, 0cm},clip]{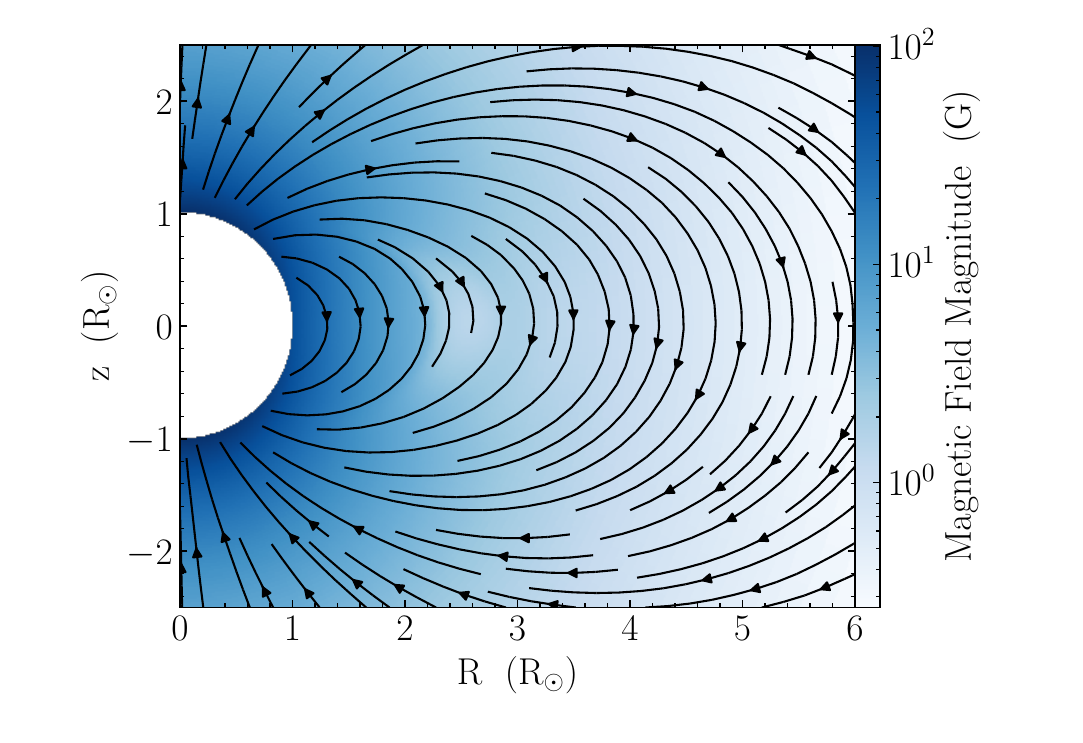}
	\caption[]{Density (top left), temperature (top right), Energy difference, given by equation \ref{eq:cooling} (bottom left) and magnetic field magnitude with field line visualisation (bottom right). All profiles are at the time of a significant condensation, $\simeq~6.7$ days into the simulation. The thermalised loop, with footpoints at $\theta_{\mathrm{l}}$ and $\theta_{\mathrm{r}}$ is seen as the smaller loop within the closed magnetosphere. A large fraction of the loop has condensed into a cool, high density structure that is undergoing collapse, moving down to the southern hemisphere footpoint of the loop, where it is absorbed by the chromosphere. The energy difference is negative in the higher density regions, indicating that cooling due to radiative losses is dominant over heating. An video version of this figure is provided online \label{fig:condens_event}}
\end{figure*}

\begin{figure*}
	\centering	
	\includegraphics[width=0.99\textwidth,trim={0cm, 0cm, 0cm, 0cm},clip]{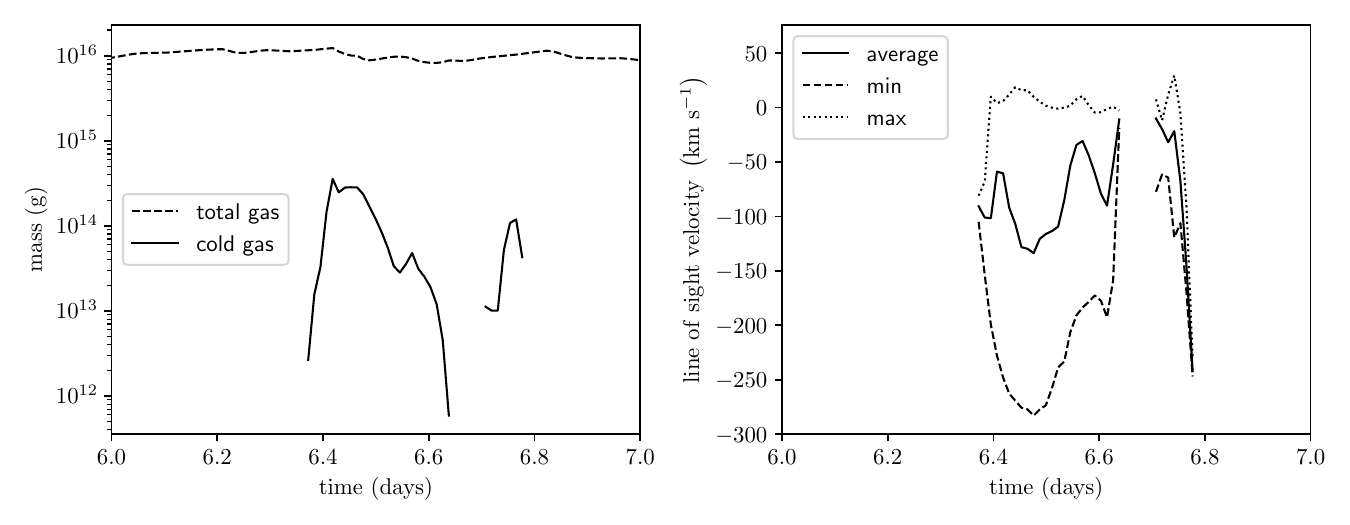}
	\caption[]{Time series of the gas in the simulation that has a temperature $<80\times10^{3} \ \mathrm{K}$. Below this cutoff, we assume the gas has condensed and has a significant population of neutrals. Gaps in the time series indicate a lack of material below the temperature cutoff. Left: time series of the mass in the simulation within the source surface (assumed to be $2.5 \ R_{\ast}$ based on the results from Fig. \ref{fig:spectrum}). The peak cold gas mass is $3.6\times10^{14} \ \mathrm{g}$. Right: time series of the line-of-sight velocity for the condensation event shown in Fig. \ref{fig:condens_event}, for an observer aligned with the stellar equator. The maximum (dotted line), minimum (dashed line) and mass weighted (solid line) velocity show there is a spread in velocities both towards and away from the observer and that the bulk motion is towards the stellar surface. 
 \label{fig:mass}}
\end{figure*}

Fig. \ref{fig:condens_event} captures a large condensation event mid way through the 3rd stage of the TNE cycle. This structure is shown in a snapshot in Fig. \ref{fig:condens_event} and captures an in-progress condensation occurring at $\sim~6.7$ days into the simulation. The cold dense gas is not stationary, but begins to collapse while forming an arch in Fig. \ref{fig:radial_mass}.

Fig. \ref{fig:mass} shows a detailed time series of this event. We find that the structure has a peak mass of $3.6 \times 10^{14}$ g, corresponding to a maximum coronal cold gas component of 2.9\%. The feature has a spread of line-of-sight velocities between $50 \ \mathrm{km} \ \mathrm{s}^{-1}$ (blue shifted) and $-250 \ \mathrm{km} \ \mathrm{s}^{-1}$ (red shifted) and lasts for $\simeq~5$ hr. This time period is shorter than any of the peak periods found in Fig. \ref{fig:spectrum}. The data in Fig. \ref{fig:condens_event} is produced via a temperature cut at $80 \times 10^{3}\ \mathrm{K}$. This allows us to capture gas that has condensed and excludes most of the mass of the hot corona. From the velocity spread, we can see that at the time when the condensation cools below our temperature cut, it is already traveling at approximately $100 \ \mathrm{km} \ \mathrm{s}^{-1}$ (red shifted). This implies that we are only capturing the latter half of the falling gas trajectory, happening within a narrow window (see the arcs in Fig. \ref{fig:radial_mass}). This implies that the majority of the $37 \ \mathrm{hr}$ time period for the large arcs is comprised of the rise and concentration of hot gas before the cool phase, which occurs in the fall to the chromosphere.

We might expect that once a condensation forms, it will fall back to the surface in approximately the free-fall time, which for a height of 1 $R_{\ast}$ above the stellar surface is $t_{ff}~\approx~1.12 \ \mathrm{hr}$. This is shorter by approximately a factor of four than the life time of the condensation event ($\simeq~5$ hr). If we calculate the terminal velocity from $t_{ff}$ we obtain a value of $170 \ \mathrm{km} \ \mathrm{s}^{-1}$ (red shifted) which is consistent with our simulations. This suggests that the magnetic geometry of our thermalised loop is playing an important role in supporting the condensed gas.  The loop does not drain to the chromosphere in the manner in which one discreet object would fall under gravity, but instead as an elongated structure spanning the majority of the loop. This way, there can be gas draining at speeds close to the free fall velocity over a period many times greater than the free fall time. We note that although the free-fall speed is not a strong function of stellar mass (since it depends on the ratio of stellar mass to radius), we would expect the free-fall time to be shorter for M dwarfs because of their smaller size.

We stop short of calling this feature a prominence as the condensed gas is neither stable nor stationary within the thermalised loop. Even as the structure begins to form it collapses towards the surface with a spread of velocities. It is more appropriate to call it large-scale coronal rain. This is consistent with the notion that our simulated star has a dynamical magnetosphere, where material is free to move between the chromosphere and the closed corona in the TNE cycle, rather than a centrifugal magnetosphere where it is suspended in the coronal by rotational acceleration. In our current setup, the magnetic geometry does not lead to a stable point for a prominence to build. 

What distinguishes these two types of magnetospheres is the relative position of the Kepler and Alfv\'{e}n radii \citep{Petit2012}, which in our study are $R_{\rm{K}} = 4.2 R_{\ast}$ and $R_{\rm{A}} \simeq 41 R_{\ast}$ (average) respectively. As $R_{\rm{K}} < R_{\rm{A}}$, our simulated star should be strongly in the centrifugal magnetosphere regime. However, from the dynamics of what we observe this is not the case. To address this discrepancy, we consider that the relative position of $R_{\rm{K}}$ and $R_{\rm{A}}$ as a characterising factor, stems from the analysis of massive star magnetospheres. This may not be appropriate in the case of cool stars. From Fig. \ref{fig:radial_mass}, we can see that the extent of the closed magnetosphere determines the height to which material can be supported or confined by magnetic tension. This is given by the source surface, $R_{\rm{SS}}\simeq 2.5 R_{\ast}$ and not $R_{\rm{A}}$. If we consider $R_{\rm{SS}}$ instead of $R_{\rm{A}}$ we can indeed see that $R_{\rm{SS}} < R_{\rm{K}}$ and that we are in the dynamical magnetosphere regime.






The range of velocities exhibited by this condensation event is consistent with observations of other stars, where velocities range from the fairly low values of $70 \ \mathrm{km} \ \mathrm{s}^{-1}$ (red shifted) detected in the Ultracool Dwarf VB10 \citep{2022ApJ...925..155K}, to the ranges of $100-200  \ \mathrm{km} \ \mathrm{s}^{-1}$ (red shifted) reported by \citet{2022ApJ...928..180W} for an M dwarf. These latter values are typical of those found in M dwarfs:  \citet{2023ApJ...945...61N} found $200-500  \ \mathrm{km} \ \mathrm{s}^{-1}$ (red shifted) for YZCMi and \citet{2019A&A...623A..49V} found in a survey of 25 M dwarfs that $100-300  \ \mathrm{km} \ \mathrm{s}^{-1}$ (blue and red shifted) was a typical range. These values may be below the escape speeds, but faster condensations have also been observed, such as the detection of a fast-moving condensation in AD Leo of $5800 \ \mathrm{km} \ \mathrm{s}^{-1}$ (blue shifted) \citep{1990A&A...238..249H}. The number of such observations is not yet sufficient to detect a clear trend with stellar mass, but larger velocities have been found on more massive stars, such as $510 \ \mathrm{km} \ \mathrm{s}^{-1}$ (blue shifted) on EK Dra \citep{2022ApJ...926L...5N} or $760-1690 \ \mathrm{km} \ \mathrm{s}^{-1}$ (blue shifted) on the RS CVn-type star V1355 Orionis \citep{2023ApJ...948....9I}. 

Values for of velocities for solar coronal rain are, however, typically much lower. \citet{2022FrASS...920116A} found a range of $100-150  \ \mathrm{km} \ \mathrm{s}^{-1}$ (red shifted), while \citet{2023arXiv230508775S} found $\simeq 40 \ \mathrm{km} \ \mathrm{s}^{-1}$ (red shifted). Sun-as-a-star measurements also tend to give lower values. \citet{2022ApJ...933..209N} found $95 \ \mathrm{km} \ \mathrm{s}^{-1}$ (red shifted) while  \citet{2022ApJ...939...98O} reported velocities up to $200 \ \mathrm{km} \ \mathrm{s}^{-1}$ (red shifted).

While the velocities of our large condensation event may be consistent with stellar observations, the masses we derive are smaller. This one event has a peak mass of 3.6 $\times$ 10$^{14}$g which is below the lower end of reported masses from the sample of M dwarfs considered by Vida  ($10^{15} - 10^{18}$g) or indeed the values of $10^{18}$g reported for another M dwarf by \citet{2022ApJ...928..180W} or 7.7 $\times$ $10^{17}$g reported for AD Leo \citep{1990A&A...238..249H}. A slightly larger mass (1.1 $\times$ $10^{18}$g) is reported for the G-type star EK Dra  \citep{2022ApJ...926L...5N}. The values for the RS CVn-type star V1355 Orionis are higher (9.5 $\times$ $10^{18}$g $<$ M $<$ 1.4 $\times$ $ 10^{21}$g) \citep{2023ApJ...948....9I}. 

We can only speculate on the origin of this discrepancy. Such small masses for our simulated condensation (and their correspondingly small H$\alpha$ transients) may simply have escaped observational detection.  A calculation of these H$\alpha$ signatures is complicated by the fact that any condensation may have been heated as it moves through the hot corona, in the same way that solar ejected prominences may suffer heating \citep{2016A&A...589A.128H}. Indeed, based on NLTE modelling of the H$\alpha$ transients on the M dwarf V374 Peg, \citet{2022MNRAS.513.6058L} concluded that they may be as hot as 95,000-115,000 K. In the case of this cooler star, there is a further complicating factor in that the lower stellar surface brightness may mean that the thermal contribution of the source function dominates over pure scattering, with the result that the transient may be seen in emission rather than absorption, even as it transits the star.  Even for the solar-mass star that we have modelled, however, a calculation of the mass and the H$\alpha$ signature of any condensation is limited by the 2D nature of our model, where the mass we analyse is effectively a thin segment of the full magnetosphere. The impact of this will be tested in future fully 3D simulations.

\section{Conclusions}

We have analysed the structure and dynamics of the coronae of rapidly-rotating solar-like stars through 2.5D MHD simulations. We find that large-scale condensations form naturally in response to excess heating. These form at heights $1~-~2 \ R_{\ast}$ above the stellar surface - much further out in the corona than similar solar condensations, despite the fact that the extent of the closed-field corona is similar to that of the Sun. These condensations display complex velocity structures, with some upflows, but predominantly downflows at velocities similar to those of solar coronal rain. The range of velocities is large - extending to around $250 \ \mathrm{km} \ \mathrm{s}^{-1}$ - and similar to many of the velocities seen in observations of stellar H$\alpha$ line asymmetries.
We find that these cold clumps comprise some $\sim3\%$ of the entire coronal mass. The peak mass of one of the cold clumps is small by stellar standards at $3.9\times10^{14}\ \mathrm{g}$, much less than the $10^{16} - 10^{17}$g inferred from the H$\alpha$ line asymmetries in the M-dwarf sample of \cite{2019A&A...623A..49V} or of other more massive stars. 

These condensations display quasi-periodic behaviour, with dominant periods around 37 hours coexisting with shorter periods. These shorter periods are typical of those detected in the coronal rain that forms on much smaller scales on the Sun  e.g. 9, 5.6 and 3.8 hrs reported by \citet{2015ApJ...807..158F} with shorter periods of order 30 mins reported by \citet{2022ApJ...931L..27S}.

Our condensation events mirror closely those seen in simulations of solar loops, either at similar scales or at larger scales. Therefore we demonstrate that the same physics occurs on cool-stars other than the Sun and that the young Sun, with a faster rotation rate and magnetic field strength may have experienced the same coronal rain phenomenon.

\section*{Data availability}
Data is available upon request to the corresponding author.

\section*{Acknowledgements}

The authors thank the anonymous reviewer for their helpful comments and suggestions; which improved the quality and content of the publication.

The authors thank C. Downs and R. Keppens for instructive conversations that improved the physics and methods used in this publication. We also acknowledge technical support from R. Beckmann, without which this publication would not be possible. SD-Y and MJ acknowledge support from STFC consolidated grant number ST/R000824/1. CDJ acknowledges support from the NASA GSFC Internal Scientist Funding Model (competitive work package) program. This research was supported by the International Space Science Institute (ISSI) in Bern, through ISSI International Team project 545 (``Observe Local Think Global: What Solar Observations can Teach us about Multiphase Plasmas across Physical Scales'').

\section*{ORICD iDs}

Simon Daley-Yates \orcidlink{0000-0002-0461-3029} \url{https://orcid.org/0000-0002-0461-3029} \\
Moira M. Jardine \orcidlink{0000-0002-1466-5236} \url{https://orcid.org/0000-0002-1466-5236} \\
Craig D. Johnston \orcidlink{0000-0003-4023-9887} \url{https://orcid.org/0000-0003-4023-9887}

\bibliographystyle{mnras}
\bibliography{references}

\label{lastpage}

\end{document}